\documentclass[twocolumn]{emulateapj}
\usepackage{graphicx}
\usepackage{dcolumn}
\usepackage{color}
\usepackage{hyperref}
\usepackage{amsmath}

\def\arcsec{\hbox{$^{\prime\prime}$}}
\def\deg{\hbox{$^\circ$}}
\def\init{\hspace{0.75 mm}}
\def\gal{J1329+3234}

\begin{document}

\title{An Optically Obscured AGN in a Low Mass, Irregular Dwarf Galaxy: A Multi-Wavelength Analysis of \gal}

\author{N.\init J.\init Secrest\altaffilmark{1}, S.\init Satyapal\altaffilmark{1}, M.\init Gliozzi\altaffilmark{1}, B.\init Rothberg\altaffilmark{1,2,3}, S.\init L.\init Ellison\altaffilmark{4}, W.\init S.\init Mowry\altaffilmark{1}, J.\init L.\init Rosenberg\altaffilmark{1}, J.~Fischer\altaffilmark{5}, and H.\init Schmitt\altaffilmark{5} }

\altaffiltext{1}{George Mason University, Department of Physics \& Astronomy, MS 3F3, 4400 University Drive, Fairfax, VA 22030, USA}

\altaffiltext{2}{Leibniz-Institut f\"{u}r Astrophysik Potsdam (AIP), An der Sternwarte 16, D-14482, Potsdam, Germany}

\altaffiltext{3}{LBT Observatory, University of Arizona, 933 N.~Cherry Ave., Tuscan, AZ 85721, USA}

\altaffiltext{4}{Department of Physics and Astronomy, University of Victoria, Victoria, BC V8P 1A1, Canada}

\altaffiltext{5}{Naval Research Laboratory, Remote Sensing Division, 4555 Overlook Ave SW, Washington, DC 20375, USA}

\begin{abstract}
Supermassive black holes are found ubiquitously in large, bulge-dominated galaxies throughout the local Universe, yet little is known about their presence and properties in bulgeless and low mass galaxies.  This is a significant deficiency, since the mass distribution and occupation fraction of non-stellar black holes provide important observational constraints on supermassive black hole seed formation theories and many dwarf galaxies have not undergone major mergers that would erase information on their original black hole population.  Using data from the \textit{Wide-field Infrared Survey Explorer}, we discovered hundreds of bulgeless and dwarf galaxies that display mid-infrared signatures of extremely hot dust highly suggestive of powerful accreting massive black holes, despite having no signatures of black hole activity at optical wavelengths.  Here we report, in our first follow-up X-ray investigation of this population, that the irregular dwarf galaxy J132932.41+323417.0 ($z=0.0156$) contains a hard, unresolved X-ray source detected by \textit{XMM-Newton} with luminosity  $L_\textrm{2-10~keV}=2.4\times10^{40}$~erg~s$^{-1}$, over two orders of magnitude greater than that expected from star formation, strongly suggestive of the presence of an accreting massive black hole.  While enhanced X-ray emission and hot dust can be produced in extremely low metallicity environments, J132932.41+323417.0 is not extremely metal poor ($\approx$ 40\% solar).  With a stellar mass of~$2.0\times10^8$~$M_\sun$, this galaxy is similar in mass to the Small Magellanic Cloud, and is one of the lowest mass galaxies with evidence for a massive nuclear black hole currently known.

\end{abstract}

\keywords{Galaxies: active --- Galaxies: dwarf --- X-ray: Galaxies  --- Infrared: Galaxies --- Black hole physics}

\section{Introduction}
\label{intro}

It is well known that supermassive black holes (SMBHs) are ubiquitous in massive bulge dominated galaxies. In contrast, for decades, less than a handful of SMBHs were known in either low mass, or bulgeless galaxies \citep[e.g.,][]{filippenko2003,barth2004}.  Identifying such a population and constraining their occupation fraction is crucial to our understanding of the origins of SMBHs and the secular pathways for their growth \citep[e.g.,][]{ volonteri2010,vanwassenhove2010} . In recent years, there have been a number of discoveries of active SMBHs in the low bulge mass regime \citep[e.g.,][]{dewangan2005, satyapal2007, satyapal2008, satyapal2009, shields2008, ghosh2008, izotov2008, desroches2009, gliozzi2009,mcalpine2011, reines2011, reines2012, secrest2012, secrest2013, dong2012, araya2012, reines2013, coelho2013, bizzocchi2014, satyapal2014, reines2014, maksym2014}. While the list of active galactic nuclei (AGNs) in low mass or bulgeless galaxies may be steadily growing, optically identified AGNs likely represent a small fraction of the total number of AGNs in this population. AGNs identified by their optical spectra are preferentially found in massive bulge-dominated hosts, with the fraction of galaxies with optical signs of accretion dropping dramatically at stellar masses $\log{M_\star/M_\sun}<10$~\citep[e.g.][]{kauffmann2003}.  For example,~\citet{reines2013} find that out of 25,974 low-mass ($\log{M_\star /M_\sun}<$9.5) galaxies with high signal-to-noise optical spectra, only 151 show any optical signatures of AGN activity (0.6\%) and only 35 (0.1\%) are unambiguously identified as AGNs on the BPT diagram.  While a combination of obscuration, energetically weak AGNs, and contamination by star formation all limit optical AGN selection in late-type galaxies, optical AGN selection in lower mass galaxies may also be affected by their lower metallicities, reducing their [N~II]/H$\alpha$ emission line ratios and shifting AGN-hosting galaxies into the HII region of the BPT diagram~\citep[see][and references therein]{izotov2008}, and even unobscured AGNs may nonetheless fail to be discerned optically~\citep[e.g.,][]{pons2014}.  The use of optical AGN diagnostics in low mass galaxies is further complicated in practice by the non-uniform nature of SDSS spectroscopic targeting, which preferentially targets more luminous galaxies~\citep[i.e., not dwarfs; see also the discussion in][]{greene2007a}. 

Motivated by the possibility that optical studies miss a significant fraction of AGNs in bulgeless and low mass galaxies, we used the the all-sky \textit{Wide-field Infrared Survey Explorer} (\textit{WISE}) to search for optically hidden AGNs in a large sample of bulgeless galaxies \citep[]{satyapal2014}. Remarkably, we discovered several hundred optically normal bulgeless galaxies, many with low stellar masses, that display extreme red mid-infrared colors ([3.4$\micron$]-[4.6$\micron$]; hereafter $W1$-$W2$) suggestive of dominant AGNs that may outnumber optically-identified AGNs by as much as a factor of $\approx6$. Astonishingly, the vast majority of these galaxies have no signs of accretion at optical wavelengths. Moreover, based on their red \textit{WISE} color, not only do these galaxies possibly harbor AGNs, but the AGNs must dominate the mid-IR luminosities of their galaxies, since in typical low redshift sources, it is very difficult to replicate the observed \textit{WISE} colors from star formation alone~\citep[e.g.,][]{assef2013}.

While the red mid-IR colors discovered by \textit{WISE} are highly suggestive of accretion activity, there are cases when red $W1$-$W2$ colors indicate hot dust due to extreme star formation, especially when associated with red $W2$-$W3$ ([4.6$\micron$]-[12$\micron$]) colors. Indeed, there have been a handful of low metallicity blue compact dwarfs (BCDs) with extreme mid-infrared colors \citep[e.g.,][]{griffith2011,izotov2014} raising the possibility that there is a similar origin for such mid-IR colors in bulgeless galaxies. It is impossible to confirm the putative AGN in this population of galaxies with the infrared observations alone. Follow up X-ray observations are required, since the detection of a hard, luminous nuclear X-ray source coincident with the nucleus provides unambiguous confirmation of the existence of accreting black holes.

In this paper, we present the first follow-up X-ray observations obtained with \textit{XMM-Newton} of the bulgeless galaxy SDSS~J132932.41+323417.0 (SDSS ``objectID''=1237665126939754596; hereafter \gal). \gal~is a blue, irregular dwarf galaxy ($\log{M_\star /M_\sun}$=8.3) with a metallicity of $Z/Z_\sun$=0.4, at a redshift of $z$=0.0156 that hosts a bright, unresolved mid-IR source with \textit{WISE} colors suggestive of the presence of an AGN ($W1$-$W2$=0.73~mag.; $W2$-$W3$=2.59~mag.). Based on the optical emission line ratios from the SDSS spectrum, there is no evidence for an AGN ($\log{\rm{[OIII]_{\lambda5007}/H\beta}}$ = 0.25, $\log{\rm{[NII]_{\lambda6584}/H\alpha}}$ = $-1.05$).

We adopt a standard $\Lambda$CDM cosmology with $H_0=70$~km~s$^{-1}$~Mpc$^{-1}$, $\Omega_{\textrm{M}}=0.3$, and $\Omega_\Lambda=0.7$.

\section{Observations and Data Reduction}
\subsection{XMM Observations}
We obtained X-ray observations of \gal~on 2013 June 12, as part of a broader \textit{XMM-Newton} program searching for AGN X-ray signatures in bulgeless galaxies (Secrest~et~al.~2015,~\textit{in prep.}). We acquired three full-frame event files with the EPIC pn, MOS1, and MOS2 CCDs, in an effective energy range of 0.3-10 keV, with effective exposure times of 30.0 ks for pn and 34.8 ks for the two MOS detectors.  We reduced the data with the \textit{XMM-Newton Science Analysis Software} (SAS), version 13.0.0, using the most up to date calibration files. We checked our reduced event files for hard X-ray background flares ($>10$ keV), but no significant flaring occurred during the exposure.  

Of the 135 EPIC sources detected in the 30$\arcmin$ field of view of our observation, we found one source coincident with \gal~(hereafter S1) and one source $\sim20\arcsec$ to the SW of \gal~(hereafter S2) on-axis with an angular resolution of 6$\arcsec$.  In Figure~\ref{astrometry}, we show the 2-10 keV pn image of \gal. Neither source shows any evidence of being extended, as confirmed by the automated \textit{XMM-Newton} pipeline.

\subsection{Large Binocular Telescope Observations}
Since there are two X-ray sources detected in our \textit{XMM-Newton} observations, we obtained an optical spectrum of our target that includes S1 and S2 on UT Feb 17, 2014 using the Multi-Object Double Spectrograph 1~\citep[MODS-1;][]{pogge2006,pogge2010} mounted on one of the two 11.8 meter mirrors of the Large Binocular Telescope (LBT).  MODS-1 is a low-medium resolution two channel (blue \& red) spectrograph mounted on the f/15 Gregorian focus of the SX (left) mirror of the LBT. In two channel mode MODS-1 can simultaneously observe the wavelength range 0.3-1.0\micron.  The observations were obtained using only the red arm of MODS-1, which covers 0.49-1.00\micron. The spatial scale of the red arm is $0{\arcsec}\hspace{-1mm}.123$~pixel$^{-1}$.  A $1{\arcsec}\hspace{-1mm}.2$ wide segmented longslit mask was used.  The mask contains 5 longslit segments, each 60{\arcsec} in length.  A single segment was sufficient to span S1 and S2.  The observations were obtained under thin clouds and in near full moon conditions, for a total of 2400 seconds (2x1200 second exposures).  The spectrophotometric standard Feige 67 was observed using the same instrumental setup as our science observations.  Slitless pixel flats were obtained to correct for the detector response. Ar, Xe, Kr, Ne and Hg lamps were observed for wavelength calibration and to measure the spectral resolution.  The slit was centered on S2 with RA=13\textsuperscript{h}29\textsuperscript{m}31\textsuperscript{s}\hspace{-0.5mm}.269, Decl.=+32\deg33\arcmin59\arcsec\hspace{-0.5mm}.91, and a position angle of 227\deg.  Note that while the slit covers much of \gal, we did not center it on \gal, as we did not want to our optical spectrum to overlap with the spectrum from SDSS.  Instead, we oriented the slit such that it would cover the K-band concentration possibly associated with S1 in Figure~\ref{ukidssk} (see \S\ref{NIRsrc1}).

The data were first reduced using Version 0.3 of the modsCCDRed collection of Python scripts. These perform basic 2D reduction including: bias subtraction, removing the overscan regions, constructing flat fields from the slitless pixel flats, fixing bad columns and flipping the red arm data so that wavelength increases from left to right along the x-axis.  Further processing was performed using customized \textsc{iraf} scripts developed by B.\init R.  Cosmic rays were removed using the task \texttt{CRUTIL}.  Two dimensional spectra were extracted in strip mode for the central longslit segment using the task \texttt{APALL}.  MODS-1 spectra are tilted along both the spectral (x-axis) and spatial (x-axis) dimensions.  The spectra for both exposures were simultaneously corrected in both axes using the arc lines and the \textsc{iraf} tasks \texttt{ID}, \texttt{REID}, \texttt{FITC}, and \texttt{TRANSFORM}.  MODS-1 data are very sensitive to the polynomial order used to wavelength calibrate the data. A 4th order Legendre polynomial produces the smallest residuals and avoids introducing low-order noise into the 2D spectra.  From the arc lamps, the dispersion was measured to be 0.84~{\AA}~pixel$^{-1}$.  Measurements of the arc lines yielded a final spectral resolution of 6.77{\AA} for the $1{\arcsec}\hspace{-1mm}.2$ slitwidth. This corresponds to $R\sim$740-1380 over the observed wavelength range.  Once the 2D spectra were rectified and wavelength calibrated, the background was subtracted in each exposure by fitting a 3rd order Chebyshev polynomial to the columns (spatial axis) using the \textsc{iraf} task \texttt{BACKGROUND}.  The two exposures were then combined and flux calibrated using the spectro-photometric standard Feige 67. This step also removes the instrumental signature.  

Finally, the data were corrected for atmospheric extinction and galactic reddening assuming $R_\textrm{V}=3.1$ and a value of $A_\textrm{V}=0.020$~\citep{schlafly2011}.  A 1D spectra of S1 was extracted from the combined 2D rectified and background subtracted image in an aperture of width 100 pixels or $12{\arcsec}\hspace{-1mm}.3$ to maximize signal-to-noise, using the task \texttt{APALL}.  A 1D spectra of S2 was extracted in the same fashion, but using an aperture of width 10 pixels or $1{\arcsec}\hspace{-1mm}.23$. The 1D spectra were then flux calibrated (removing the instrumental response), corrected for atmospheric extinction, and corrected for galactic reddening using the same values as above.

\subsection{Astrometry}

\begin{figure}
\noindent{\includegraphics[width=8.7cm]{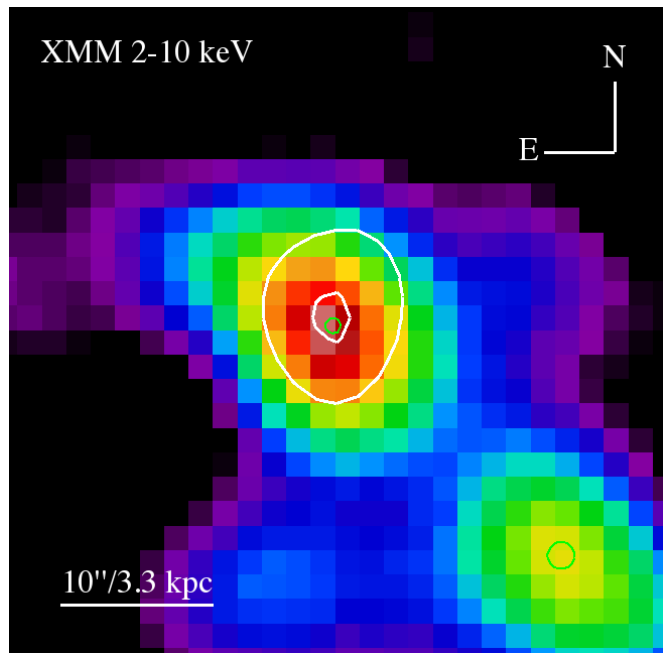}}
\caption{Smoothed pn 2-10 keV image with \textit{WISE} $4.6\micron$ source contours overlaid in white.  The pn image was binned by a factor of 32 and smoothed with a Gaussian kernel of 4 pixels.  The green circles represent the centroids of S1 to the NE and S2 to the SW, with radii representing their $1\sigma$ formal astrometric uncertainties.\\}
\label{astrometry}
\end{figure}

The EPIC X-ray event files for \gal~were not astrometrically corrected. This introduced a $\sim4\arcsec$ offset between the location of nearby X-ray sources (including S2) and their putative optical counterparts in the SDSS field.  We manually inspected the \textit{XMM-Newton} pipeline astrometric solution of the optical monitor $B$-band image by blinking it with the SDSS $g$ image, and found it to be sufficient to apply to our X-ray event files.  This astrometric solution applies an offset of $\Delta$RA=$-3.2\arcsec$ and $\Delta$Decl.=$3.4\arcsec$.  The astrometry corrected EPIC data confirms that the position of S1 does coincide with the position of the red \textit{WISE} source (Figure~\ref{astrometry}).  The \textit{XMM-Newton} pipeline formal astrometric uncertainties (the astrometric uncertainty \textit{after} correction) is 0.47$\arcsec$ and 0.51$\arcsec$ for S1 and S2, respectively.

\begin{figure}
\noindent{\includegraphics[width=8.7cm]{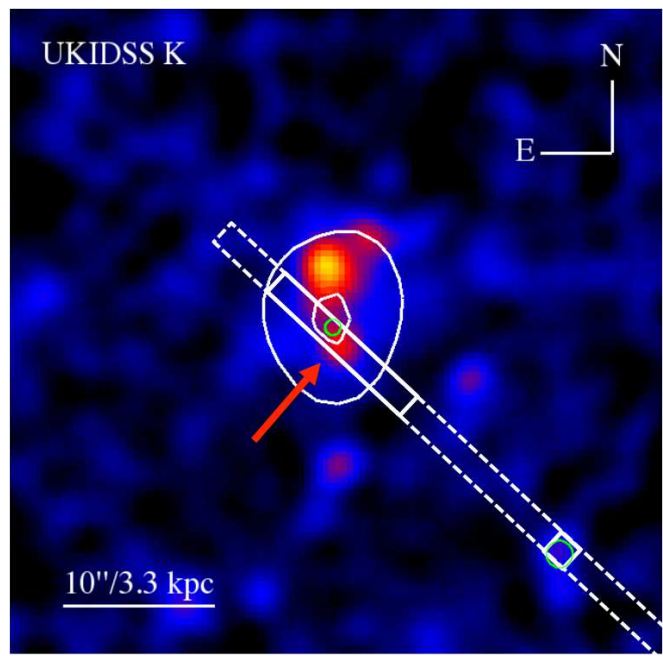}}
\caption{UKIDSS $K$ band image of \gal~with the (unresolved) \textit{WISE} $4.6\micron$ source overlaid as contours.  The UKIDSS and \textit{WISE} astrometric accuracies are $0.1\arcsec$ and $0.2\arcsec$, respectively.  The green circles represent the centroids of S1 to the NE and S2 to the SW, with radii representing their $1\sigma$ formal astrometric uncertainties.  The position of the LBT slit is outlined in dashed line, with the apertures used for S1 and S2 solid ($12.3\arcsec$ and $1.2\arcsec$, respectively).  The red arrow marks the possible near-IR counterpart to S1.\\}
\label{ukidssk}
\end{figure}

\begin{figure}
\noindent{\includegraphics[width=8.7cm]{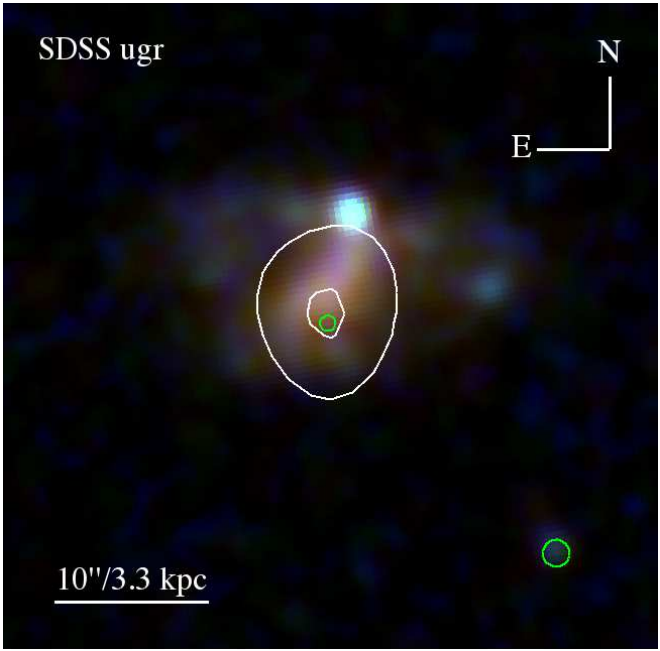}}
\caption{Smoothed SDSS $ugr$ image of \gal~with \textit{WISE} $4.6\micron$ source contours overlaid.  The green circles represent the centroids of S1 to the NE and S2 to the SW, with radii representing their $1\sigma$ formal astrometric uncertainties.\\}
\label{SDSSugrWISEcont}
\end{figure}

While \gal~is not bright enough to have been detected by 2MASS, it was detected in the UKIDSS survey\footnote{\url{www.ukidss.org/}}, enabling us to examine the near-infrared $K$ band (Figure~\ref{ukidssk}).  \gal~in the $K$ band displays an irregular, clumpy structure, with no discernible photo center. Optically, \gal~also presents an irregular structure, with no unambiguous photo center (Figure~\ref{SDSSugrWISEcont}).  The location of S2 is consistent with an optical concentration with no obvious association with \gal, and S2 likely a background AGN (see \S\ref{src2discuss}).  This optical concentration source does not appear in any QSO catalogs and does not appear to have been studied before in any way.

\subsection{X-ray Analysis}

We extracted X-ray spectra for both sources following standard procedures, and grouping the spectra by a minimum of 15 counts per channel for the $\chi^2$ statistic.  Due to the projected proximity of the sources ($\sim20\arcsec$), we extracted spectra using apertures that corresponded to the half-energy diameter of an on-axis point-source ($\sim15\arcsec$ for pn) in order to minimize cross-contamination.  We performed spectral analysis on the binned spectra using \textsc{xspec} version 12.7.0~\citep{arnaud1996} and using the online \textit{Chandra} Colden\footnote{\url{cxc.harvard.edu/toolkit/colden.jsp}} tool to find the Galactic neutral hydrogen column density for \gal, which we account for in our spectral analysis as a fixed photo-electric absorption multiplicative component (\texttt{wabs}).  We do not explicitly state the inclusion of this spectral model later in the paper, because it is included in all our X-ray spectral modeling, and has a value of $N_\textrm{H}=1.08\times10^{20}$~cm$^{-2}$.  While we primarily focused our spectral analysis on the higher signal-to-noise pn data, we also performed simultaneous spectral fitting on the data from all three detectors by multiplying each spectrum by a separate constant that was free to vary, holding the values of the spectral parameters identical between spectra.  This allowed inter-detector sensitivity variability to be accounted for without compromising the overall spectral fit.  The value of this constant was generally close to unity.  To derive fluxes and errors, we appended the \texttt{cflux} convolution model, holding the model normalization fixed.  All errors in this work are $1\sigma$.

\section{Results}

\subsection{S1}
\label{src1results}

\begin{figure}
\noindent{\includegraphics[width=8.7cm]{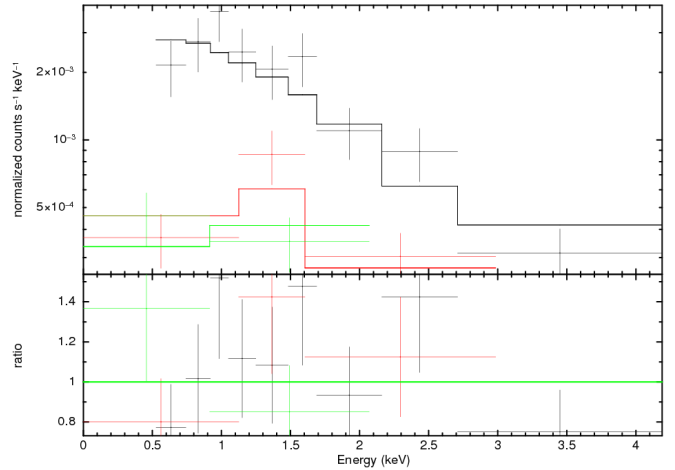}}
\caption{Best fit spectral model for S1.  The black line represents the pn data and the green and red line represent the data from the two MOS detectors.\\}
\label{Source1spec}
\end{figure}

We fit the X-ray spectrum for S1 with a simple power-law spectrum, achieving a good fit ($\chi^2$/degrees of freedom (d.o.f.) = 11/10=1.1) with $\Gamma=1.3\pm0.1$.  We achieve a comparable fit with the addition of an intrinsic absorber (\texttt{zwabs}, $\chi^2$/d.o.f. = 8.6/9 = 0.96), which yields an intrinsic absorption of $N_\mathrm{H} = (1.4\pm0.1)\times10^{21}$~cm$^{-2}$ and a power law index of $\Gamma=1.8\pm0.4$, but we do not find the fit to be sensitive to this addition due to low counts ($F$-test probability $\sim$30\%; pn, MOS1, and MOS2 counts = $149\pm13$, $49\pm7$, and $42\pm7$ counts, respectively).  In Figure ~\ref{Source1spec}, we show the best-fit spectrum for S1. We derive unabsorbed full and hard band X-ray fluxes of $\log{F_\textrm{0.3-10~keV}}=-13.3\pm0.1$~erg~cm$^{-2}$~s$^{-1}$ and $\log{F_\textrm{2-10~keV}}=-13.4\pm0.1$~erg~cm$^{-2}$~s$^{-1}$, translating to a full band luminosity of $L_\textrm{0.3-10~keV}=(3.0\pm0.7)\times10^{40}$~erg~s$^{-1}$ and a hard X-ray luminosity of $L_\textrm{2-10~keV}=(2.4\pm0.6)\times10^{40}$~erg~s$^{-1}$, similar to the hard X-ray luminosities found in LINERS and low-luminosity Seyferts~\citep[e.g.,][]{terashima2002}.  We note that if we had used the model with the intrinsic absorber, then we would have X-ray fluxes of $\log{F_\textrm{0.3-10~keV}}=-13.3\pm0.1$~erg~cm$^{-2}$~s$^{-1}$ and $\log{F_\textrm{2-10~keV}}=-13.6\pm0.1$~erg~cm$^{-2}$~s$^{-1}$, almost identical to the simple power law model, and so the exact choice of model does not significantly affect our results.

\begin{figure}
\noindent{\includegraphics[width=8.7cm]{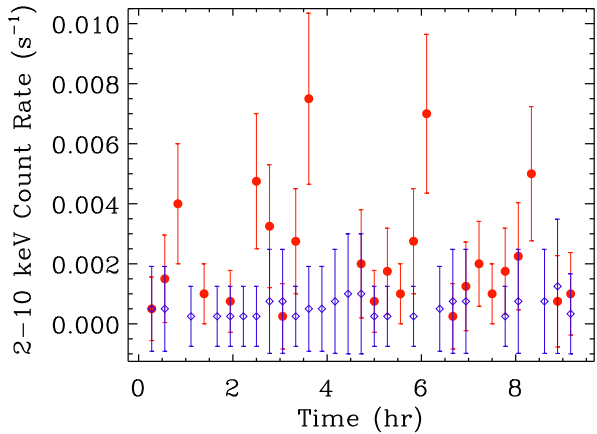}}
\caption{\textit{Red:} Hard X-ray light curve of S1.~\textit{Blue:} Background hard X-ray light curve.\\}
\label{lc-hard}
\end{figure}

Given the vicinity of S2 to S1, we extracted the S1 light curves from a circular region with a radius of $11\arcsec$ and the background light curve from a larger nearby region without X-ray sources. We extracted light curves in the soft (0.3-2 keV), hard (2-10 keV), and total (0.3-10 keV) energy bands, and tried different time bins (from 250s to 1000s) to investigate the presence of variability based on a $\chi^2$ test. Only the hard light curves shows evidence for variability irrespective of the time bin used. However, a closer look at the hard background-subtracted light curves reveals the presence of spurious negative values (they are about one order of magnitude smaller that the average count rate), which can be explained by statistical fluctuations of the background light curve combined with the low count rate of the source. Once these spurious data points are removed, only the 1000s light curve shows some suggestive evidence for variability. Specifically, using the ``lcstats'' routine from \texttt{Xronos}\footnote{\url{https://heasarc.gsfc.nasa.gov/xanadu/xronos/xronos.html}} we got the following results: for the 2-10 keV light curve with 25 data points $\chi^2$=34.7 with a corresponding probability of constancy $P_\chi^2$=0.07; for the 0.3-10 keV light curve with 34 data points $\chi^2$=33.9 and $P_\chi^2$=0.42; whereas for the soft 0.3-2 keV light curve with 33 data points $\chi^2$=35.8 and $P_\chi^2$=0.29. We therefore conclude that only the hard light curve is not consistent with the hypothesis of constancy. In Figure~\ref{lc-hard}, we show the 2-10 keV background-subtracted light curve (filled red circles) along with the background light curve (open diamonds, which does not show any significant variability: $\chi^2$=16 for 25 degrees of freedom and $P_\chi^2$=0.91) where the time bins are 1000s.

\subsubsection{Near-IR Counterpart of S1}
\label{NIRsrc1}

While there is no obvious optical counterpart for S1, the UKIDSS dataset reveals a possible counterpart in the near-IR.  The source, marked with a red arrow in Figure~\ref{ukidssk}, is unresolved (median seeing $\sim0.5\arcsec$\footnote{\url{www.ukidss.org/sciencecase/proposal/proposal.wfcam.html}}), and corresponds to ``sourceID''=433793068380 in the UKIDSSDR9PLUS database, at RA=13\textsuperscript{h}29\textsuperscript{m}32\textsuperscript{s}\hspace{-0.5mm}.39, Decl.=+32\deg34$\arcmin$13$\arcsec$\hspace{-0.5mm}.1.  The near-IR $Y$, $H$, and $K$ magnitudes for this source are $20.3\pm0.2$, $18.5\pm0.1$, and $17.0\pm0.1$, respectively.  The $J$ band magnitude was not measured successfully.  This source possibly corresponds to SDSS ``objectID''=1237665126939754597, which has a very faint $r$ band magnitude of $23.2\pm0.3$, but this source's photometry suffers from flux interpolation and deblending problems, so this magnitude value should be considered tentative.

\subsubsection{Optical Spectrum of S1}

\begin{figure}
\noindent{\includegraphics[width=8.7cm]{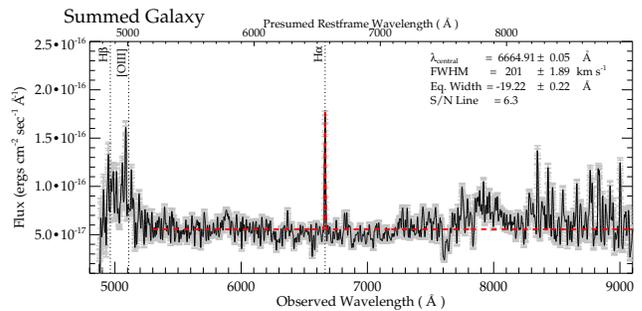}}
\caption{The LBT optical spectrum of the S1 counterpart.  The red line denotes the fit to the continuum and H$\alpha$ line, while the gray denotes the RMS noise. The dotted lines on the left denote where H$\beta$ and [O~III]5007{\AA} would be if they were detected. The apparent structure between 7600{\AA} and 8200{\AA} is due to noise and is not real. At $\lambda$ $>$ 8100{\AA} the night-sky lines for these observations subtracted out rather poorly, yielding noisy data.\\}
\label{Source1opticalspec}
\end{figure}

The position of the LBT slit overlapped with the position of the near-IR counterpart to S1 described above, as well as the position of S2.  In Figure~\ref{Source1opticalspec}, we plot the 1D optical spectrum corresponding to the position of this near-IR counterpart.  While we detect an H$\alpha$ line at $\lambda$=6665.5{\AA} ($z$=0.0156), we do not detect any other lines at a significant level.  The RMS for the H$\beta$ and [O~III]5007{\AA} lines is $\approx8\times10^{-18}$~erg~cm$^{-2}$~s$^{-1}$ and $\approx7\times10^{-18}$~erg~cm$^{-2}$~s$^{-1}$ for the [N~II]6584{\AA} line.  The total H$\alpha$ flux within the 12.3$\arcsec$ aperture is $F_\mathrm{H\alpha}$=$(5.92\pm0.37)\times10^{-16}$~erg~cm$^{-2}$~s$^{-1}$.  The optical spectrum for this region therefore shows no signatures of an AGN, as optically-identified AGNs typically have [N~II]/H$\alpha$ line ratios between $\approx$ 0.6-1.2.

\subsection{S2}

\begin{figure}
\noindent{\includegraphics[width=8.7cm]{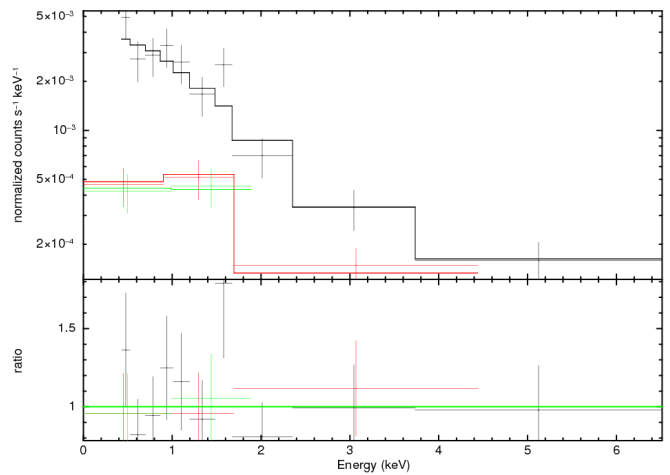}}
\caption{Best fit spectral model for S2.  The black line represents the pn data and the green and red line represent the data from the two MOS detectors.\\}
\label{Source2spec}
\end{figure}

We fit the X-ray spectrum for S2 using all three detectors, and found that the source can best be fit by a simple power law with $\Gamma=1.6\pm0.1$ (Figure~\ref{Source2spec}), although the spectra are still slightly over-fit even with a simple power law ($\chi^2$/d.o.f.=6.4/11=0.6).  We derive unabsorbed full and hard band X-ray fluxes of $\log{F_\textrm{0.3-10~keV}}=-13.4\pm0.1$~[erg~cm$^{-2}$~s$^{-1}$] and $\log{F_\textrm{2.0-10~keV}}=-13.6\pm0.1$~[erg~cm$^{-2}$~s$^{-1}$], translating to full and hard X-ray luminosities of $L_\textrm{0.3-10~keV}=(1.2\pm0.3)\times10^{42}$~erg~s$^{-1}$ and $L_\textrm{2-10~keV}=(7.3\pm1.7)\times10^{41}$~erg~s$^{-1}$ given its redshift~(See~\S\ref{src2discuss}). We do not find evidence for variability in any energy band.

\subsubsection{Optical Spectrum of S2}
\label{src2discuss}
\begin{figure}
\noindent{\includegraphics[width=8.7cm]{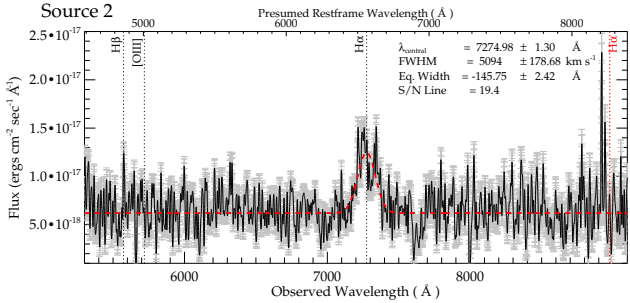}}
\caption{LBT Optical Spectrum of S2.  The dotted line on the far right shows where H$\alpha$ would be if the broadened line detected were H$\beta$. The dotted lines on the left denote where H$\beta$ and [O~III]5007{\AA} would be if they were detected, and the gray shading denotes the RMS noise.\\}
\label{Source2opticalspec}
\end{figure}

Figure~\ref{Source2opticalspec} shows the final 1D spectrum corresponding to S2 covering the wavelength range from 5300-9100{\AA}.  At $\lambda$ $>$ 8100{\AA} the night-sky lines for these observations subtracted out rather poorly, yielding noisy data (see for example the noise spike at 8930{\AA} in Figure~\ref{Source2opticalspec}).  Only one emission line is detected in the spectrum at $6.7\sigma$. Although a confirmed redshift cannot be ascertained from this data, we can constrain a likely redshift based on the assumption that the detected line is H$\alpha$. Given the broad wavelength range shown, if the line were H$\beta$, then we would see H$\alpha$ in the same window, assuming Case B recombination \citep{osterbrock89}, at least 2.86 times as strong as H$\beta$.  We can also eliminate the possibility that the line is Lyman-$\alpha$ because there is strong continuum detected significantly blueward of the expected Lyman limit; there are no strong emission lines corresponding to CIV or CIII (all of which would fall within the wavelength range of our observations), and the flux associated with both the line and the \textit{XMM-Newton} observations would be uncharacteristically bright for the required redshift of $z\sim5$.  The absence of H$\beta$, or rather, the failure to detect H$\beta$ is not unexpected if the H$\alpha$ from S2 arises from a Broad Line Region (BLR) in a radio galaxy or Seyfert 1.8/1.9, where the Balmer decrement can be up to a factor of 10~\citep[e.g.,][]{osterbrock76,osterbrock81,crenshaw88}. If the object is a QSO or Seyfert 1 the decrement could be up to a factor of 5~\citep[e.g.,][]{osterbrock77,dong2005}.  All of these scenarios would place the H$\beta$ line beyond the detection limits of the MODS-1 observations.  Thus, assuming the detected emission line is H$\alpha$, S2 would therefore be a background AGN at $z=$0.108 and not associated with \gal.  The total H$\alpha$ flux within the 1.2$\arcsec$ aperture is $F_\mathrm{H\alpha}$=$(8.1\pm1.2)\times10^{-16}$~erg~cm$^{-2}$~s$^{-1}$.

\section{Discussion}
 
\subsection{X-ray Emission from X-ray Binaries}
\label{xraybinaries}

Since the X-ray luminosity of \gal~is low compared to powerful AGNs, we investigated the possibility that the observed X-ray emission is due to the integrated emission from a population of low mass and high mass X-ray binaries (LMXRBs and HMXRBs).  Numerous studies of galaxies without AGNs have show that non-AGN X-ray emission is a function of the star formation rate (SFR) and the stellar mass of the host galaxy~\citep[e.g.,][]{ranalli2003,boroson2011,lehmer2010}.  Because of the large difference between the SDSS fiber area and the X-ray extraction aperture (7 arcsec$^2$ vs.~180 arcsec$^2$), applying a simple geometric aperture correction to the extinction-corrected H$\alpha$ luminosity introduces too much uncertainty into the estimate for the star formation rate.  We therefore use the SFR from the Max Planck Institut fur Astrophysik/Johns Hopkins University (MPIA/JHU) collaboration\footnote{\url{www.mpa-garching.mpg.de/SDSS/}}, which follows~\citet{brinchmann2004} with photometric corrections following~\citet{salim2007}.  In short, for galaxies without optical line emission evidence for AGN activity, fiber SFRs are ``aperture corrected'' by fitting stellar population models to photometry outside of the fiber and then calculating the galaxy-wide SFRs.

\begin{figure}
\noindent{\includegraphics[width=8.7cm]{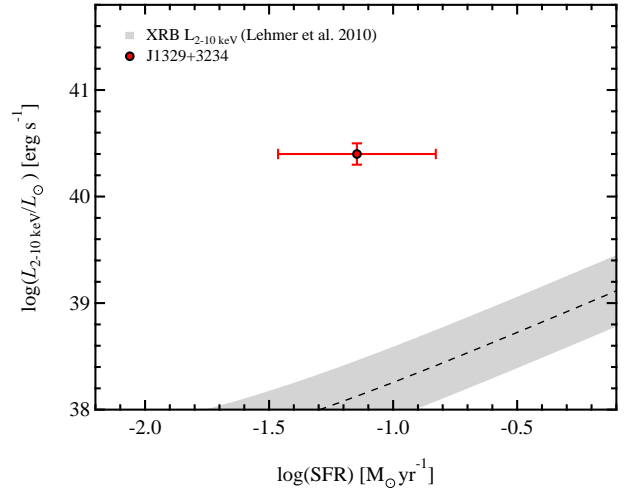}}
\caption{X-ray luminosity expected from the combined emission from X-ray binaries as a function of star formation rate from \citet{lehmer2010} together with the observed $L_\textrm{2-10~keV}$ luminosity of \gal.  The gray shading indicates the $1\sigma$ scatter on the \citet{lehmer2010} relation (0.34 dex).\\}
\label{lehmer}
\end{figure}

We note, however, that this represents the full, galaxy-wide SFR, and so somewhat overestimates any possible X-ray binary contamination within the \textit{XMM-Newton} point-spread function.  To remain as conservative as possible, we nonetheless use this SFR, which is 0.07~$M_\sun$~yr$^{-1}$.  Using the findings of~\citet{lehmer2010}, this translates to a hard X-ray luminosity due to X-ray binaries of $L_\textrm{2-10~keV}=1.3\times10^{38}$~erg~s$^{-1}$, with a $3\sigma$ upper limit of $L_\textrm{2-10~keV}=1.4\times10^{39}$~erg~s$^{-1}$, still well over an order of magnitude less than what we observe (see Figure~\ref{lehmer}).

It is important to note, however, that the galaxies used in the sample of~\citet{lehmer2010} were massive, luminous infrared galaxies (LIRGs), and so it may not be the case that the calibration factors presented therein are accurate for a dwarf galaxy such as \gal.  Indeed, recent studies \citep[e.g.,][]{mapelli2010,kaaret2011, prestwich2013} have reported enhanced production of X-ray emission from XRBs in extremely low-metallicity galaxies relative to the star formation rate, presumably as a result of the presence of more massive stars and an enhancement of the accretion rates in low metallicity environments. While \gal~is not an extremely metal poor dwarf galaxy for its mass ($\approx$ 40\% solar) we nonetheless estimated the X-ray emission produced by star formation assuming the relation found in extremely metal poor dwarfs.  Using a sample of BCDs with $Z/Z_\sun<0.1$, \citet{brorby2014} find that the X-ray emission produced for a given SFR is approximately an order of magnitude larger than that found in near solar metallicity galaxies~\citep[see also][]{kaaret2011}. Even if we combine this calibration factor with the $3\sigma$ upper limit to the hard X-ray luminosity due to the galaxy-wide SFR, the X-ray emission predicted from star formation ($L_\textrm{2-10~keV}=1.4\times10^{40}$~erg~s$^{-1}$) is still about half the value that we observe.

As a further check on the possibility of HMXRB contamination, we explored the possibility that there may be heavily obscured star formation that would not reveal itself using traditional optical diagnostics.  To obtain an extinction-insensitive upper limit on the SFR, we used the source lower sensitivity limit at 1.4 GHz derived from the NRAO VLA Sky Survey (NVSS\footnote{The National Radio Astronomy Observatory is a facility of the National Science Foundation operated under cooperative agreement by Associated Universities, Inc.}) survey, in which \gal~is not detected, which has a typical lower flux density limit for detected sources of 2.5~mJy.  We derive a non-detection 1.4 GHz luminosity upper limit of $L_\textrm{1.4~GHz}<1.5\times10^{21}$~W~Hz$^{-1}$.  Combining this luminosity limit with Eq.~23 from~\citet{condon1992} for the 1.4 GHz thermal free-free emission due to star formation, we derive an upper limit to the star formation rate of SFR($M$~$>$$~5~M_\sun$)~$<2.8$~$M_\sun$~yr$^{-1}$.  Propagating this through the~\citet{lehmer2010} relation, we find an upper limit to the HMXRB X-ray luminosity of $\log{L_\textrm{2-10~keV}}<39.7$, still nearly an order of magnitude less than what we measure, and less than what we measure at the $+2\sigma$ level on the~\citet{lehmer2010} relation.  Furthermore, the angular resolution limit of the NVSS survey is $\sim45\arcsec$, encompassing all of \gal~and is considerably more than the angular resolution of \textit{XMM-Newton}, and so this upper limit on the SFR should be considered very conservative.

Finally, we note that the likely variability of S1 (See \S\ref{src1results}), indicates that the X-ray emission is originating from a single source, further making it unlikely that a population of XRBs is responsible for the emission.
 
\subsection{ULX Origin for the X-ray emission}
\label{ulxorigin}

Given the relatively low X-ray luminosity of S1, we investigated the possibility that the X-ray source in \gal~is an ultraluminous X-ray source (ULX). ULXs are off-nuclear X-ray sources with luminosities in excess of 10$^{39}$~erg~cm$^{-1}$, which is the Eddington luminosity of a $10~M_\sun$ stellar mass black hole. The luminosities of ULXs can be produced either by anisotropic emission (beaming) or super Eddington accretion from a stellar sized black hole or by accretion onto intermediate mass black holes (IMBHs), the latter scenario still being the subject of significant controversy~\citep[for a recent review see][]{feng2011}. ULXs are generally rare; however they are preferentially found in regions of enhanced star formation~\citep[e.g.,][]{gao2003, mapelli2008} and low metallicity~\citep[e.g.,][]{prestwich2013, thuan2014}, and the occurrence rate per unit galaxy mass is higher in dwarfs than more massive galaxies~\citep{swartz2008}. Recently ~\citet{somers2013} found 16 ULXs in 8 bulgeless galaxies, some with luminosities in excess of $10^{40}$~erg~cm$^{-1}$.  This raises the possibility that the X-ray source in \gal~is a ULX, and therefore possibly consistent with a stellar sized black hole.

We investigated the possibility that the X-ray source in \gal~is a ULX by comparing the X-ray and mid-infrared properties of \gal~with those of definitive AGNs and ULXs that are cleanly separated from their galaxies' nucleus. In AGNs, the hard X-ray emission and the mid-infrared continuum emission is strongly correlated \citep{lutz2004}.  This is expected because the hard X-ray emission provides a direct view of the central engine, and the nuclear infrared continuum is dominated by thermally reprocessed radiation from the AGN.  We derived our sample of AGNs in the following manner: We cross-matched the final \textit{WISE} all-sky data release catalog (AllWISE\footnote{\url{wise2.ipac.caltech.edu/docs/release/allsky/}}) with the 3XMM-DR4 catalog to within less than $<1\arcsec$, and obtained redshifts from SDSS DR10\footnote{\url{www.sdss3.org/dr10/}} in a similar manner.  We required that the \textit{WISE} $W1$, $W2$, and $W3$ band fluxes have a signal-to-noise ratio greater than or equal to $\geq5$, and that there be no photometric quality or source extent flags.  We similarly required that the \textit{XMM-Newton} hard band ``SC\_EP4\_FLUX'' and ``SC\_EP5\_FLUX'' values have a combined signal-to-noise ratio greater than or equal to $\geq5$, and that there be no high background level (flaring) flags.  We further required that the X-ray sources show no sign of spatial extent by requiring the ``SC\_EXTENT'' flag be equal to 0, and that the source observation off-axis angles be less than $<5\arcmin$ to minimize off-axis aberrations.  We finally required redshifts greater than $z\geq0.01$ to reduce distance errors.  This selection yielded 184 AGNs with 2-10 keV luminosities between $\log{L_\textrm{2-10~keV}}$ = 42.1-46.7 and $W2$ luminosities between $\log{L_{W2}}$ = 42.2-46.5 [erg~s$^{-1}$].  The mean mid-IR colors of these 184 AGNs are $W1$-$W2$ = 1.01 mag.~and $W2$-$W3$ = 2.98 mag., with standard deviations of 0.29 mag. and 0.35 mag., respectively.  We estimated the linear regression fit between $L_\textrm{2-10~keV}$ and $L_{W2}$ by drawing $10^5$ random samples with replacement (bootstrapping) of our 184 AGNs, finding a strong correlation (Pearson $r=0.91$) with:
$$
\log{L_\textrm{2-10~keV}}= (0.93\pm0.03)\cdot\log{L_{W2}}+(3.26\pm1.25)
$$
with a standard deviation of $\sigma$ = 0.40~dex in the $\log{L_\textrm{2-10~keV}}$ direction and the units of luminosity used being erg~s$^{-1}$.  For comparison, we fit the 6$\micron$ continuum/2-10 keV X-ray luminosity data used for a sample of Sy 1 and Sy 2 galaxies in~\citet{lutz2004}, and found good agreement, with a slope of $0.95\pm0.08$, intercept of $1.63\pm3.50$, and $\sigma$=0.43~dex.

\begin{figure}
\noindent{\includegraphics[width=8.7cm]{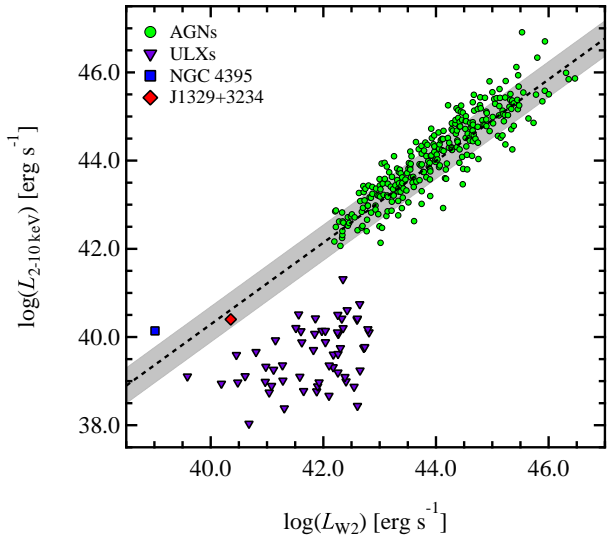}}
\caption{The observed 2-10~keV luminosity versus the W2 luminosity for our sample of AGNs, ULXs, and \gal.  For comparison, we also plot NGC~4395, the archetypical bulgeless dwarf Seyfert 1 galaxy.  The dashed line represents the linear regression described in \S\ref{ulxorigin} with the $1\sigma$ scatter in gray.\\}
\label{lxlw2plot}
\end{figure}

In order to compare the mid-IR and X-ray properties of these AGNs with those of ULXs, we used the most recent and comprehensive catalog of ULXs by \citet{walton2011}, which consists of 470 ULX candidates, located in 238 nearby galaxies, generated by cross-correlating the 2XMM Serendipitous Survey \citep{watson2009} with the Third Reference Catalogue of Bright Galaxies.  We crossmatched this catalog with AllWISE to within less than $<1\arcsec$ and find that 228 ULXs are associated with \textit{WISE} sources. Of these, only 57 had \textit{WISE} photometry with $W1$, $W2$, and $W3$ detections with signal-to-noise greater than $3\sigma$, and \textit{all} are extended in the \textit{WISE} bands, suggesting that these ULXs are not being detected in the mid-IR themselves, but rather are simply associated with large-scale, extended emission from star formation. In Figure~\ref{lxlw2plot}, we plot the hard X-ray luminosity versus the $W2$ band luminosity for AGNs, ULXs, and \gal, along with the linear regression derived above.  As can be seen, \gal~ has mid-IR and X-ray properties completely consistent with AGNs, while ULXs do not.  We further note in Figure~\ref{lxlw2plot} that NGC 4395, the archetypical bulgeless dwarf Seyfert 1 galaxy, also follows this trend derived from much brighter AGNs, implying that AGNs do not deviate significantly from this relationship for a wide range of X-ray luminosities.

\begin{figure}
\noindent{\includegraphics[width=8.7cm]{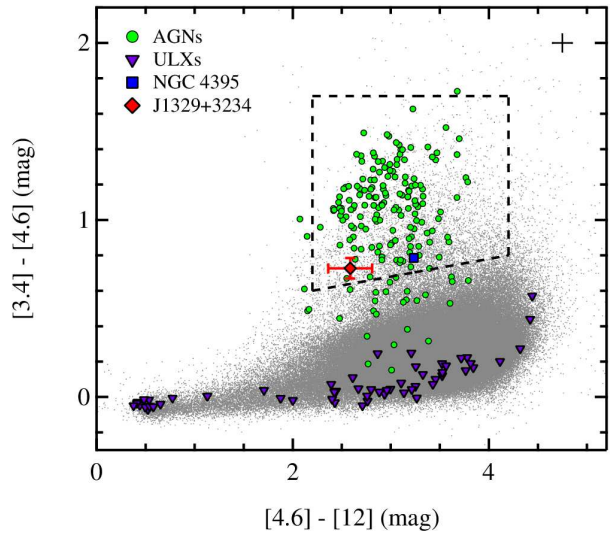}}
\caption{$W1$-$W2$ color versus the $W2$-$W3$ color for AGNs, ULXs, and \gal, with NGC 4395 included for comparison. The AGN region from \citet{jarrett2011} is also shown as the dashed line, and the typical uncertainty for the AGN colors is given by the black cross at the top right.\\}
\label{colorcolorplot}
\end{figure}

The ULXs are also associated with significantly different mid-IR colors than AGNs. The average $W1$-$W2$ color associated with these 57 ULX sources detected by \textit{WISE} is 0.07, with a maximum of 0.57.  In Figure~\ref{colorcolorplot}, we plot the $W1$-$W2$ versus $W2$-$W3$ colors of the ULXs, our AGN sample from above, \gal, normal galaxies from SDSS, and the 3-band AGN demarcation region from \citet{jarrett2011}.  As can be seen, the ULXs are clearly separated from \gal, which falls well within the AGN region of the color-color diagram, as do the majority of our sample AGNs.  Even amongst the most  extreme ULXs ($L_\textrm{2-10~keV}>10^{40}$~erg~s$^{-1}$) in this catalog \citep{sutton2012}, all of which are well separated from the galaxy nuclei and are unlikely to be background AGNs, only 3 are associated with \textit{WISE} sources, all of which are resolved and none have red \textit{WISE} colors.

\subsection{Red \textit{WISE} Colors in Low Mass Starburst Galaxies}

Given the above considerations, the only plausible explanation for the mid-IR and X-ray properties of \gal~is the presence of an AGN.  We note, however, that \gal~is certainly not the only dwarf galaxy with extreme or unusual mid-IR colors.  For example, \citet{izotov2011} find, from a sample of $\sim5000$ SDSS galaxies with \textit{WISE} colors, 4 dwarf galaxies with extreme \textit{WISE} colors $W1$-$W2>2$~mag.  Like \gal, these galaxies are in the mass range of $M_\star\sim10^{8-9}$~$M_\sun$, and like \gal, these galaxies are not particularly metal-poor dwarfs, with $Z/Z_\sun$=0.2-0.5.  However, these four galaxies occupy a completely different region of mid-IR color space than \gal, not only with $W1$-$W2$=2.13-2.37, but most notably with $W2$-$W3$=3.58-4.76, indicating considerably more emission at longer wavelengths from dust heating due to starbursts.  Indeed, the SFRs of these four galaxies are in the range of 5.8 to 26.5 $M_\sun$~yr$^{-1}$, at least two orders of magnitude higher than \gal.  Finally, there exists a single \textit{Chandra} observation of one of the four galaxies from~\citet{izotov2011}, J1457+2232, observed 2007 March 23 with ACIS-I for 7.2 ks (ObsID 7709; PI: Garmire). The galaxy is not detected, with a $3\sigma$ upper limit to the 2-10 keV flux of $\sim3\times10^{-15}$~erg~cm$^{-2}$~s$^{-1}$, assuming a similar X-ray spectrum to \gal.  Given this galaxy's relatively high mid-IR luminosity of $L_\textrm{W2}=5.0\times10^{42}$~erg~s$^{-1}$, this galaxy is at least two orders of magnitude underluminous in the hard X-rays compared to AGN-hosting galaxies (see \S\ref{ulxorigin}), with a $3\sigma$ hard X-ray upper luminosity limit of $L_\textrm{2-10~keV}<2.0\times10^{41}$~erg~s$^{-1}$.  Thus, we conclude that the colors of \gal, along with the SFR and the X-ray luminosity, are not consistent with a population of low-mass starburst galaxies.

\subsection{Black Hole Mass}

Because the ratio of the hard X-ray luminosity to the mid-IR luminosity is consistent with what is seen in AGNs, it is unlikely that we are seeing beamed X-ray emission.  With this in mind, we can immediately calculate the Eddington lower mass limit.  Using a conservative bolometric correction factor of $\kappa:=L_\textrm{bol.}/L_\textrm{2-10~keV}=15$~\citep{vasudevan2007,vasudevan2009}, we get a lower mass limit of $M_\textrm{BH}>2.9\times10^3~M_\sun$.  More realistically, the black hole is not radiating at its theoretical upper limit.  If the black hole is accreting at a rate similar to that found for the sample of low-mass black holes from~\citet{greene2007b}, then $L_\textrm{bol.}/L_\textrm{Edd.}\sim0.4$ and the black hole mass is $M_\textrm{BH}=7.3\times10^3~M_\sun$.  There is considerable spread in their sample, however, and several low-mass black holes radiate as low as $L_\textrm{bol.}/L_\textrm{2-10~keV}\sim0.02$.  In this case, the $M_\textrm{BH}$ in \gal~has a mass of about $M_\textrm{BH}\sim1.5\times10^5~M_\sun$.

\subsection{Host Galaxy Properties}

 \gal~is a very low luminosity dwarf.  The absolute $g$ band magnitude  is -16.3, almost two magnitudes fainter than the LMC~\citep{tollerud2011}, and is fainter than about $\approx98\%$ of the host galaxies in the sample of dwarfs with optical AGNs from \citet{reines2013}, which ranges from -15.2 to -21.2 (mean -18.2, $\sigma=0.8$, number of galaxies $N=151$).  These are in turn considerably fainter than previous sample of low mass optically identified AGNs from \citet{greene2007b}, which range from -17.6 to -21.4 (mean -19.7, $\sigma=0.8$, $N=229$), and the sample of \citet{dong2012}, which range from -17.4 to -22.4 (mean -20.3, $\sigma=0.9$, $N=309$).  
 
 Based on the SDSS images, the host galaxy reveals an irregular clumpy morphology in the optical, typical of irregular dwarf galaxies.  \gal~is slightly extended for a galaxy in its mass range, with a Petrosian half-light radius $r_{50}$ = 1.82~kpc, which is more extended than about $\sim75\%$ of the galaxies in the mass range $\log{M_\star /M_\sun}=8.1-8.3$, and somewhat more extended than the original definition of a blue compact dwarf galaxy (BCD; \citet{thuan1981}). We used the two dimensional parametric fitting program \textsc{galfit} \citep{peng2010} to model the $g$ band image of \gal.  The best fit model consists of a single Sersic with $n$ = 0.9, consistent with an exponential disk and two off-nuclear clumpy structures.  There is no evidence for a bulge component based on the SDSS images and no clear spiral structure apparent in the SDSS bands. While AGNs are found in disk galaxies, the fraction of optically selected AGNs even within low mass galaxies is higher in galaxies with higher Sersic indices \citep [e.g.,][]{reines2013}. The $g$-$r$ color of \gal~is -0.5~mag., comparable to the median color of the dwarfs studied by \citet{reines2013}. 
  
The stellar mass  of \gal~from the NASA-Sloan Atlas (NSA\footnote{\url{www.nsatlas.org}}; ID=93798) is $\log{M_\star /M_\sun}$ = 8.3, similar in mass to the SMC~\citep[e.g.,][]{skibba2012}, and less massive than any of the 35 optically-identified AGNs dwarf galaxies from \citet{reines2013} (lowest mass galaxy NSA ID=60536, $\log{M_\star /M_\sun}=8.4$), although similar to the lowest mass, optically normal galaxy in their sample with broad H$\alpha$ emission (NSA ID = 15952, $\log{M_\star /M_\sun}=8.4$), identified as a possible low-metallicity AGN in~\citet{izotov2007}.  If more galaxies in this mass range are confirmed to host AGNs through X-ray observations, then a key result from this study is that the fraction of AGNs at low stellar masses, revealed by \textit{WISE}, may be much higher than the fraction of AGNs revealed optically~\citep{satyapal2014}.  This is in direct contrast with results from optical spectroscopic surveys, which show that the AGN fraction approaches nearly 100\% for all emission line galaxies at the highest masses and drops dramatically with decreasing stellar mass \citep[see Figure 5, solid histogram in top panel, in][]{kauffmann2003}.

\section{Summary and Conclusions}

We have conducted our first follow-up X-ray observations of a newly discovered population of low mass and bulgleless galaxies that display extremely red mid-infrared colors highly suggestive of a dominant AGN despite having no optical signatures of accretion activity.  Our main results can be summarized as follows: \\

\begin{enumerate}
\item{Using \textit{XMM-Newton} observations, we have confirmed the presence of a hard X-ray point source consistent with AGN activity in \gal, an optically normal dwarf galaxy with red infrared colors obtained from \textit{WISE}.\\}
\item{The X-ray luminosity of \gal~is  $L_\textrm{2-10~keV}\sim2.4\times10^{40}$~erg~s$^{-1}$ . Assuming that the black hole is radiating at the Eddington limit and using a conservative bolometric correction factor of $\kappa:=15$, this corresponds to a lower limit on the black hole mass of $M_\textrm{BH}>2.9\times10^3~M_\sun$.\\}
\item{From multi-wavelength considerations, the X-ray/mid-IR activity of \gal~is consistent with the presence of an AGN, and it is unlikely that the X-ray source in \gal~is due to X-ray binary activity or is a ULX.\\}
\item{With a stellar mass of $\sim2.0\times10^8~M_\sun$, \gal~is among the lowest mass dwarf galaxies with evidence for AGN activity currently known.\\}

\end{enumerate}

\section{Acknowledgements}

We thank the anonymous referee for a very thorough and considered review that significantly improved the quality of this paper.  N.\init J.\init S.~and S.\init S.~gratefully acknowledge support by the \textit{Chandra} Guest Investigator Program under NASA Grant G01-12126X.  J.\init L.\init R.\init acknowledges support from NSF-AST 000167932.  Basic research in astronomy at the Naval Research Laboratory is funded by the Office of Naval Research.\\

\end{document}